%% file: draft.tex
\documentclass{emulateapj}
\usepackage{bm, amssymb, aas_macros}
\usepackage[colorlinks=true, citecolor=blue, urlcolor=blue,
linkcolor=blue]{hyperref}
\input{extra_preamble.tex}
\usepackage{color}

\slugcomment{ApJ, in press}

\shorttitle{The Mass of the Milky Way from Leo I's Motion}
\shortauthors{Boylan-Kolchin et al.}

\begin{document}
\title {The Space Motion of Leo I:\\ 
  The Mass of the Milky Way's Dark Matter Halo} 
\author{Michael Boylan-Kolchin\altaffilmark{1,4,5}, James
  S. Bullock\altaffilmark{1}, Sangmo Tony Sohn\altaffilmark{2}, Gurtina
  Besla\altaffilmark{3,6}, \\
\& Roeland P. van der Marel\altaffilmark{2}}
\altaffiltext{1}{Center for Cosmology, Department of Physics and Astronomy, 4129
  Reines Hall, University of California, Irvine, CA 92697, USA}
\altaffiltext{2}{Space Telescope Science Institute, 3700 San Martin Drive,
  Baltimore, MD 21218, USA} \altaffiltext{3}{Department of Astronomy, Columbia
  University, New York, NY 10027, USA} \altaffiltext{4}{email:
  \href{mailto:m.bk@uci.edu}{m.bk@uci.edu}} \altaffiltext{5}{Center for Galaxy
  Evolution fellow} \altaffiltext{6}{Hubble fellow}

\begin{abstract}
  We combine our \textit{Hubble Space Telescope} measurement of the proper
  motion of the Leo I dwarf spheroidal galaxy (presented in a companion paper)
  with the highest resolution numerical simulations of Galaxy-size dark matter
  halos in existence to constrain the mass of the Milky Way's dark matter halo
  ($\mmw$). Despite Leo I's large Galacto-centric space velocity ($200 \,\kms$)
  and distance ($261\,\kpc$), we show that it is extremely unlikely to be
  unbound if Galactic satellites are associated with dark matter substructure,
  as 99.9\% of subhalos in the simulations are bound to their host. The observed
  position and velocity of Leo I strongly disfavor a low mass Milky Way: if we
  assume that Leo I is the least bound of the Milky Way's classical satellites,
  then we find that $\mmw > 10^{12}\,\msun$ at 95\% confidence for a variety of
  Bayesian priors on $\mmw$. In lower mass halos, it is vanishingly rare to find
  subhalos at 261 kpc moving as fast as Leo I.  Should an additional classical
  satellite be found to be less bound than Leo I, this lower limit on $\mmw$
  would increase by $30\%$. Imposing a mass weighted \lcdm\ prior, we find a
  median Milky Way virial mass of $\mmw=1.6 \times 10^{12}\,\msun$, with a 90\%
  confidence interval of $[1.0-2.4]\times 10^{12}\,\msun$.  We also confirm a
  strong correlation between subhalo infall time and orbital energy in the
  simulations and show that proper motions can aid significantly in interpreting
  the infall times and orbital histories of satellites.
\end{abstract}

\keywords{galaxies: individual (Leo I) -- Galaxy: kinematics and dynamics --
  Galaxy: halo -- Galaxy: fundamental parameters -- methods: $N$-body
  simulations}

\section{Introduction}
Satellite galaxies orbiting the Milky Way (MW) have been a rich source of
astrophysical information and cosmological confusion. Individually, these
objects probe galaxy formation at the lowest masses, luminosities, and
metallicities. As a population, the Milky Way's dwarf satellites offer one of
the few existing tests of cosmology on small scales. They also provide a sample
of tracers that can be used to estimate the mass distribution of the Milky Way's
dark matter halo. Their small number, anisotropic spatial distribution, and
unknown tangential velocities are all potential sources of biases and
misunderstandings, however.

In particular, Leo I has long been known to play on outsized role when using
satellite galaxies to determine the mass of the Milky Way's dark matter halo
($\mmw$; \citealt{zaritsky1989, fich1991, kochanek1996}) because of its large
distance and high radial velocity in the Galacto-centric frame. In the analysis
of \citet{watkins2010}, which incorporates line-of-sight velocity information
for 25 Galactic satellites in determining $\mmw$, Leo I alone contributes
approximately 27\% to the mass estimate (i.e., including Leo I leads to mass
estimates that are 25-30\% larger than those that exclude Leo I).

Incorporating Leo I into an estimate of $\mmw$ generally requires the assumption
that Leo I is indeed bound. The timing argument \citep{kahn1959}, for example,
assumes the MW and Leo I to be on an elliptical orbit, with Leo I having made a
recent pericentric pass about the Milky Way. Since Leo I is, by assumption,
bound to the Milky Way in such analyses, a large Milky Way mass is required to
match Leo I's observed distance and radial velocity (e.g.,
\citealt{zaritsky1989, li2008, sohn2012a}). If Leo I is unbound, however, the
timing argument as applied to Leo I will almost certainly overestimate
$\mmw$. Leo I's extreme kinematic properties have therefore led to substantial
debate about its status as a bound satellite of the Milky Way
\citep{zaritsky1989, byrd1994, sales2007, sohn2007, mateo2008}.

Theoretical models of halo collapse provide a further basis for interpreting
satellite orbits in \lcdm. In particular, the self-similar secondary infall
model (SSIM; \citealt{fillmore1984, bertschinger1985}) has guided much of the
progress in understanding halo collapse and virialization. In the simplest
version of the SSIM -- one where only radial motions are considered -- the
matter flow around a slight overdensity in the early universe can be derived in
a straightforward manner. Shells of matter around the overdensity initially
expand with the Hubble flow. Those shells that are bound to the perturbation
eventually reach a turn-around radius and re-collapse, with more bound shells
collapsing earlier; the virial radius of a halo is approximately equal to half
of the present day turn-around radius in the SSIM.  In this model, then,
\textit{all} accreted material must be bound; otherwise, it would not collapse
onto the overdensity.

Although the physical picture of accretion in a hierarchical universe is
somewhat different from the smooth SSIM, analyses of $N$-body simulations have
shown that phase space is often structured in a very similar manner to SSIM
predictions (e.g., \citealt{ascasibar2007, diemand2008a}). With numerical
simulations, it is possible to compute the orbits of halos falling into larger
halos. Studies of large volume cosmological simulations generally find that
unbound orbits are very rare, in accordance with the expectations of the SSIM:
\citet{wetzel2011} found that less than 2\% of orbits are unbound at all
redshifts, while \citet{benson2005} found that only 0.3\% of all orbits escape
the host halo. These analyses typically adopt the point mass limit and therefore
underestimate the potential energy of the orbit coming from the mass
distribution exterior to the halos; accordingly, these unbound fractions should
be upper limits to the true unbound fraction. While the works of Benson and
Wetzel studied orbits of merging dark matter satellites across a broad spectrum
of dark matter halo masses, \citet{deason2011} and \citet{di-cintio2012} focused
on Milky Way-mass systems, again finding that the fraction of unbound satellites
is quite low. In this paper, we show that the eight highest resolution
simulations performed to date of Milky Way-size halos all have unbound fractions
of less than 1 in 1000 (0.1\%).

In accordance with predictions of the SSIM, $N$-body simulations show that halos
falling into larger halos have a characteristic orbital energy that is roughly
invariant with time: for host halos with $\mvir \sim 10^{12}\,\msun$, this
energy corresponds to an orbital velocity of $\approx 1.1 \, \vvir$ at the
virial radius, with a $1 \,\sigma$ scatter of 25\% (\citealt{wetzel2011}; see
also \citealt{benson2005, khochfar2006}). Since halo virial masses grow with
time, due to physical mass accretion or to a time-varying virial overdensity
threshold \citep{diemand2007a, diemer2012}, subhalos accreted at early times
will be somewhat more tightly bound than those accreted more recently.  Once an
infalling halo has interacted substantially with the host, its orbital energy
may decrease owing to processes such as dynamical friction; much less
frequently, it may gain orbital energy through three-body
interactions. Fluctuations in the central gravitational potential of the host
due to, e.g., virialization or mergers may also affect subhalos'
energies. Subhalos that have been part of their host for a long time will
therefore exhibit a wide range of orbital energies, whereas recently accreted
subhalos should occupy a much narrower range in energy space (and should be less
strongly bound).

This general picture has been confirmed by \citet{diemand2008a}, who show a
strong relationship between the number of orbits a subhalo has completed and its
position in radial velocity phase space for the Via Lactea (VL) simulation
\citep{diemand2007}, and by \citet{rocha2012}, who found a clear correlation
between orbital binding energy and infall time for subhalos in the Via Lactea II
simulation (VL-II; \citealt{diemand2008}). Similar relationships are generically
expected in the SSIM, although this model does not account for energy loss
processes such as dynamical friction. Note that both in the SSIM and in
cosmological simulations, satellites falling into dark matter halos for the
first time do so on bound orbits: \textit{a first infall need not, and almost
  always does not, imply an unbound orbit}. 

The three-body interactions discussed in the previous paragraphs provide a means
of boosting the velocities of some subhalos to above the local escape velocity,
however.  This possibility has received attention in the literature recently,
often in the context of exploring ``extreme'' objects such as Leo
I. \citet{sales2007} emphasized the existence of dark matter satellites that
have gained energy through three-body interactions with other, more massive
satellites inside of their host halos. \citet{ludlow2009} examined subhalo
orbits around Milky Way halos as well and showed that many satellites travel to
large apocentric distances, much larger than would be expected from their
initial orbits, after gaining energy through the aforementioned three-body
interactions during their first pericentric pass. Even after gaining energy,
these satellites typically remain formally bound to their dark matter host,
although their apocenters can (significantly) exceed their original turn-around
radius. These ``ejected'' subhalos always have very low mass at infall (0.1\% of
their host's mass or less; \citealt{ludlow2009}) and are usually found as the
companion of a more massive infalling satellite \citep{sales2007}. An additional
phenomenon specific to the environments around pairs of similarly massive halos,
such as the Local Group, is that of ``renegade'' satellites \citep{knebe2011,
  teyssier2012}, which are objects presently near one halo but that were, at
some earlier epoch, within the virialized extent of the other halo.

Although a variety of complicated orbital histories are possible for subhalos,
such orbits constitute a small minority of the satellite population around dark
matter halos. Most subhalos at large distances ($1-2$ virial radii) from their
hosts have not undergone any complex interactions with multiple other
bodies. Analytic models predict, and \lcdm\ simulations find, that many
satellites once well within a host's virial radius will spend significant time
at larger radii \citep{mamon2004, gill2005, wang2009}, a natural consequence of
the highly eccentric nature of subhalo orbits (the median apocenter to
pericenter distance ratio for subhalos is approximately 6:1;
\citealt{ghigna1998, diemand2007a}). Distant MW satellites such as Leo I may
well fall into this class of objects.

While the uncertainty surrounding Leo I's status as a bound MW satellite is a
source of frustration, this does not diminish the importance of Leo I in
establishing the dark matter distribution around the Milky Way. There are few
kinematic tracers for $D \ga 80$ kpc (see, e.g., \citealt{gnedin2010,
  deason2012a}), and deriving mass constraints from theses tracers (typically
blue horizontal branch stars) requires assumptions about their spatial
distribution and velocity anisotropies \citep{dehnen2006, deason2012a}.

In a companion paper (\citealt{sohn2012a}; hereafter, Paper I), we presented the
first determination of Leo I's transverse velocity, based on proper motion
measurements using the unique astrometric capabilities of the \textit{Hubble
  Space Telescope} (\textit{HST}).  This provides a direct probe of the mass
distribution of the Milky Way at 260 kpc, where the potential is dominated by
the dark matter halo, and therefore a new window on $\mmw$.  In this paper, we
combine the measurement of the proper motion of Leo I from Paper I with
state-of-the-art numerical simulations of the formation of Milky Way-size halos
in \lcdm\ in order to constrain the mass of the Galaxy's dark matter halo.

There is considerable disagreement in the existing literature about
$\mmw$. Direct measures have typically focused on tracers of the inner portions
of the Milky Way's dark matter halo ($d \la 80 \,\kpc$) based on measured
velocities of giant stars, with recent results ranging from $\sim [0.8-1.0]
\times 10^{12}\,\msun$ \citep{battaglia2005, xue2008} to $1.6 \times
10^{12}\,\msun$ \citep{gnedin2010}. Most recently, \citet{deason2012a} argued
that the mass of the halo within 150 kpc likely falls in the range of $[0.5-1.0]
\times 10^{12}\,\msun$.  Using radial velocities of nearby stars from the RAVE
survey to compute the local escape velocity, \citet{smith2007} estimated that
$\mmw=0.85 \times 10^{12}\,\msun$ for a standard NFW halo, or $1.42 \times
10^{12}\,\msun$ for an adiabatically contracted dark matter halo.  

Published measurements using the radial velocities of satellite galaxies give an
even wider range, from $\sim [0.8-2.5] \times 10^{12}\,\msun$, depending on the
satellites included and assumptions about the orbits of those satellites
\citep{kochanek1996, watkins2010}. \citet{boylan-kolchin2011} and
\citet{busha2011} combined analyses of large $N$-body simulations with the
\citet{kallivayalil2006} measurement of the Large Magellanic Cloud's proper
motion and obtained $\mmw \approx 2 \times 10^{12}$ and $1.2 \times
10^{12}\,\msun$, respectively. Previous determinations of $\mmw$ based on the
Leo I timing argument have ranged from $[1-3]\times 10^{12}\,\msun$
(\citealt{zaritsky1989, kochanek1996, li2008}, Paper I), while the Milky Way-M31
timing argument is also more consistent with values of $\mmw$ that are somewhat
higher than those obtained via stellar tracers \citep{van-der-marel2012}.

Indirect constraints on $\mmw$ can be obtained from combinations of
galaxy-galaxy lensing and Tully-Fisher data. \citet{dutton2010} and
\citet{reyes2011} parameterize this combination of data in terms of the typical
ratio of a galaxy's optical velocity $V_{\rm opt}$ (the circular velocity at 2.2
disk scale radii) to its virial velocity\footnote{These authors define the
  virial velocity to be the halo's circular velocity at the radius containing an
  average density of 200 times the critical density.} $V_{200c}$.  Reyes et
al. find $V_{\rm opt} =1.27\,V_{200c}$ for galaxies with stellar masses equal to
that of the Milky Way ($M_{\star, {\rm MW}} \approx 6 \times 10^{10}\,\msun$;
\citealt{flynn2006, mcmillan2011}); including scatter in concentration at fixed
mass, this becomes $V_{\rm opt} < 1.8\,V_{200c}\;(2\,\sigma)$. Using $V_{\rm
  opt, MW}=240\,\kms$ \citep{mcmillan2011, schonrich2012}, we find that Reyes
et al. predict a median value of $\mmw \approx 2.5\times10^{12}\,\msun$, with
$\mmw \ga 1.16 \times 10^{12}\,\msun$ ($2\,\sigma$). This is similar to
abundance matching's prediction of $\mmw \approx 2.5\times 10^{12}\,\msun$ for
this stellar mass \citep{guo2010, reddick2012, moster2013}.

In light of these results, it is fair to say that the mass of the Milky Way is
known to no better than a factor of two, and possibly even worse; indeed,
\citet{klypin2002} showed that a wide range of observational data can be fit
both for $\mvir=10^{12}$ and for $2\times 10^{12}$. This uncertainty is a
fundamental limitation in understanding several pressing questions related to
galaxy formation and cosmology.  For example, it is frequently assumed that
galaxies have available to them $(\Omega_b/\Omega_m) \times \mvir$ in
baryons. If $\mmw=5\times 10^{11}\,\msun$, then essentially all of those baryons
are accounted for by observed stars and gas in and around the Galaxy; if
$\mmw=2.5 \times 10^{12}\,\msun$, then a large majority of the Milky Way's
baryons are missing \citep{fukugita2004, anderson2010, gupta2012,
  fang2013}. Another example is the predicted abundance of dark matter
satellites, which scales nearly linearly with dark matter halo mass; the
expected mass of a galaxy's most massive satellite scales in the same manner. As
a result, our interpretation of the missing satellites problem
\citep{klypin1999, moore1999} and related questions about the Galaxy's satellite
population depends on $\mmw$ \citep{boylan-kolchin2011a, boylan-kolchin2012,
  wang2012, vera-ciro2013}.

A measurement of the mass distribution of the Milky Way at large Galacto-centric
distance $(D > 100 \,{\rm kpc})$ would bring significantly more clarity to each
of these issues. Here, we use the results of Paper I in conjunction with
cosmological simulations of Milky Way-size dark matter halos to provide a
constraint on $\mmw$ based on the proper motion of Leo I ($D=260.6\,\kpc$).
This paper is structured as follows.  Section~\ref{sec:methods} describes the
simulations we employ and provides a summary of the results of Paper I. We then
compare the simulations to the radial velocity (Section~\ref{subsec:vr}) and
space velocity (Section~\ref{subsec:vtot}) of Leo I. This allows us to constrain
the Milky Way's virial mass (Section~\ref{sec:mass}). We discuss our results in
the context of expected infall times of \lcdm\ satellites in general, and Leo I
in particular, in Section~\ref{sec:implications}, and conclude with a brief
summary and discussion of our results in Section~\ref{sec:conclusion}.

\section{Simulations and Data}
\label{sec:methods}
\subsection{Simulations}
Our primary set of Milky Way-size halos are those of the Aquarius simulation
\citep{springel2008}. The simulations, and the properties of the halos, have
been extensively discussed elsewhere (e.g., \citealt{springel2008,
  springel2008a, navarro2010}), and we refer the reader to those papers for
details related to the simulations. We use the six halos (denoted A-F) simulated
at ``level two'' resolution of $\approx 10^{4}\,\msun$ per particle with
Plummer-equivalent gravitational softening length of $66\,{\rm pc}$. The
Aquarius suite was run assuming a WMAP-1 cosmology with parameters
$\Omega_m=0.25, \, \Omega_{\Lambda}=0.75, \, \sigma_8=0.9, \, H_0=73\,{\rm
  km\,s^{-1}\,Mpc^{-1}}, \,n_s=1.0$. More recent measurements prefer slightly
different values of these parameters, but these differences are not expected to
affect the assembly and internal dynamics of Milky Way-mass halos at a
substantive level and are therefore unimportant for our analysis. In
Appendix~\ref{sec:appendix}, we compare results from similarly well-resolved
halos -- VL-II and GHALO
\citep{stadel2009} -- using WMAP3 parameters to show this explicitly.

The virial mass $\mvir$ of a dark matter halo is not uniquely defined, but
rather depends on the density threshold used to define a dark matter
halo. Throughout this work, we will define $\mvir$ to be the mass within a
sphere, centered on the halo in question, containing an average density $\Delta$
times the critical density of the Universe; we use $\Delta=\Delta_{\rm vir}$,
the value derived from the spherical top-hat collapse model \citep{gunn1972,
  bryan1998}, which gives $\Delta_{\rm vir} \approx 94$ at $z=0$ for the
cosmology of the Aquarius simulations. Note that other common choices of
$\Delta$ are 200 and $200\,\Omega_m(z)$. The Aquarius halos have $0.95 <
\mvir/10^{12}\,\msun < 2.2$, which is a reasonable approximation of current
constraints on the value of the Milky Way's virial mass (see
\citealt{boylan-kolchin2012} for a recent compilation).

In the \lcdm\ cosmogony, all satellite galaxies are expected to initially form
within their own dark matter halo, outside of their host's virial radius. The
luminous satellite galaxies of the Milky Way are therefore expected to be
represented by some subset of dark matter subhalos surviving to $z=0$. The
precise relationship between satellite galaxies and dark matter subhalos is
presently unclear; however, all models of galaxy formation predict a good
correspondence between the most luminous satellite galaxies and the most massive
subhalos for some suitable definition of mass, most often the maximum mass or
circular velocity that the subhalo has ever had (e.g., \citealt{kravtsov2004,
  conroy2006, guo2010, yang2012, reddick2012, behroozi2012b, moster2013}).  We
therefore record the redshift at which a subhalo's bound mass is maximized, which
we hereafter refer to as the infall redshift\footnote{This definition of infall
  time is similar to the time a subhalo last crossed the physical boundary of
  its host but does not rely on any specific definition of the host's
  ``edge''. } $\zacc$, as well as the maximum circular velocity at that time,
$\vacc$.

Dark matter subhalos with maximum circular velocities in excess of 5 $\kms$ and
having $\vmax > 0.044\,\vvir$ (corresponding to $5 \,\kms$ for the lowest mass
halo, Aq-B) are selected for further analysis. \textit{All} halos and subhalos
satisfying this $\vmax$ threshold within one Megaparsec of each Aquarius main
halo are considered, with no additional velocity-based criteria. This ensures we
do not introduce any bias against high velocity objects in our analysis. All
distances and velocities discussed below are computed in the halo-centric frame.

\subsection{Observational Data}
\label{subsec:observations}
The observational data used in our analysis are detailed in Paper I. For the
purposes of this work, the important quantities are as follows.
\begin{itemize}
\item The Galacto-centric distance to Leo I ($D_{\rm Leo I}$):
\begin{equation} 
D_{\rm Leo I} = 260.6 \pm 13.3 \;\kpc
\end{equation}

\item The Galacto-centric radial velocity of Leo I ($V_{\rm r, Leo I}$):
\begin{equation} 
V_{\rm r, Leo I} = 167.9 \pm 2.8 \;\kms
\end{equation}

\item The Galacto-centric transverse (tangential) velocity of Leo I ($V_{\rm t,
    Leo I}$):
\begin{equation} 
V_{\rm t, Leo I} = 101.0 \pm 34.4 \;\kms\,.
\end{equation}
\end{itemize}
The measured three-dimensional
Galacto-centric space velocity of Leo I is
\begin{equation}
  \label{eq:vtot}
  V_{\rm Leo I} =199.8^{+21.8\,(+47.0)}_{-17.2\,(-29.3)}\, \kms\,,
\end{equation}
where the errors -- computed from Monte Carlo sampling of the proper motion
error space -- represent 68.27\% (95.45\%) confidence intervals about the stated
median.

\section{Interpreting the motion of Leo I}
\label{sec:velocities}
\subsection{The radial velocity of Leo I}
\label{subsec:vr}
\begin{figure}
 \centering
 \includegraphics[scale=0.55]{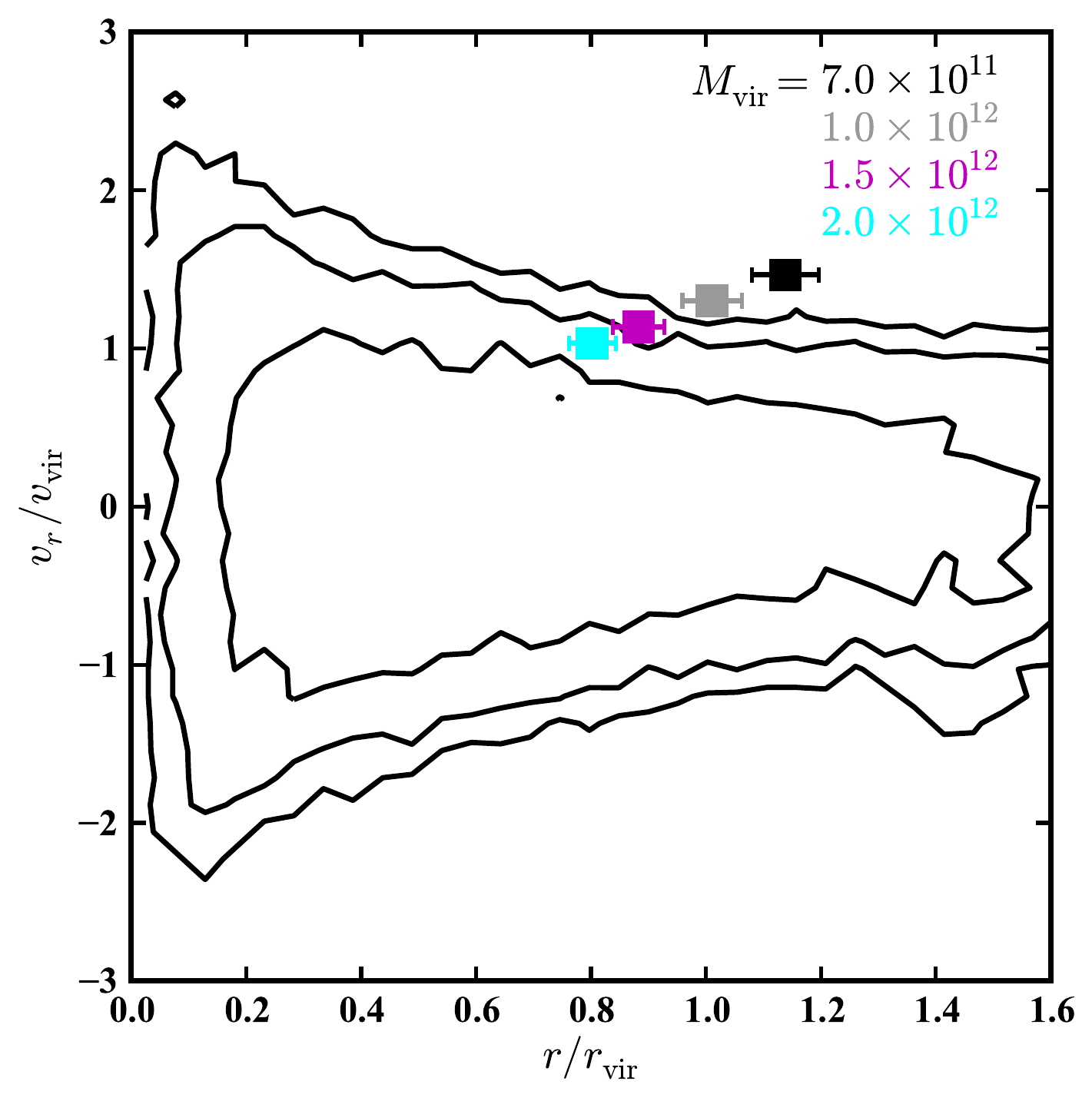}
 \caption{Radial phase space diagram of Aquarius subhalos, scaled to virial
   units. The inner, middle, and outer contours contain 68, 95.4, and 99.7\% of
   subhalos from the Aquarius simulations. Colored squares with error bars show
   the location in phase space of Leo I, with $1\,\sigma$ uncertainties, for a
   variety of assumed virial masses for the Milky Way. Note that the absolute
   error on the radial velocity is smaller than the size of the symbols. The
   observed values of $(r, \,v_r)$ for Leo I are consistent with fewer than 0.3\%
   of the Aquarius subhalos if the MW virial mass is less than $1.1\times
   10^{12} \,\msun$.
 \label{fig:vr}
}
\end{figure}
Figure~\ref{fig:vr} shows a phase space diagram for all subhalos in the Aquarius
simulations. The radial velocity $V_r$ of each subhalo is scaled by the virial
velocity of its host, and the halo-centric radius of each subhalo is likewise
scaled by the virial radius of the host. Infalling subhalos have $V_r < 0$,
while outgoing subhalos have positive radial velocities.  The black contours
contain 68.3\%, 95.5\%, and 99.7\% of subhalos\footnote{The contour value at a
  specific radius for a given percentage value is computed by finding the
  location of that value in the cumulative distribution function of $V_r$ at
  that radius.}.  Additionally, we have placed square symbols with error
bars\footnote{Unless otherwise noted, all plotted error bars are $1\,\sigma$.}
on the plot to denote the observed position of Leo I in this phase space for
four representative values of the Milky Way's virial mass:
$\mmw/10^{12}\,\msun=$ 0.7 (black), 1.0 (gray), 1.5 (magenta), and 2.0 (cyan).

It is obvious why the observed radial velocity of Leo I has demanded a high
virial mass for the Milky Way: subhalos in both the $7\times 10^{11}$ and
$10^{12}\,\msun$ halos have radial velocities as high as Leo I less than 0.3\%
of the time. In more massive halos, Leo I's high radial velocity becomes more
likely; however, it is still higher than 70\% of subhalos at Leo I's
Galacto-centric distance in a halo with $\mvir=2 \times 10^{12}\,\msun$. With
only a measurement of Leo I's radial velocity, the preferred mass of the MW is
significantly in excess of $10^{12}\,\msun$. Constraining $\mvir$ more tightly
requires knowledge of the transverse velocity of Leo I, however, as it requires
understanding how likely it is for a subhalo to be on an unbound orbit about its
host, and whether the observed tangential velocity of Leo I is cosmologically
plausible, given its radial velocity. Furthermore, as we demonstrate in
Section~\ref{sec:implications}, satellites that have been recently accreted occupy
a wide range of radial velocities but are confined to a much narrower range of
total velocities. In combination with reasonable estimates of infall times
(based on, e.g., star formation histories), 3D velocity information can
therefore be much more constraining than radial velocity information alone.

\subsection{The space velocity of Leo I}
\label{subsec:vtot}
\begin{figure}
 \centering
 \includegraphics[scale=0.55]{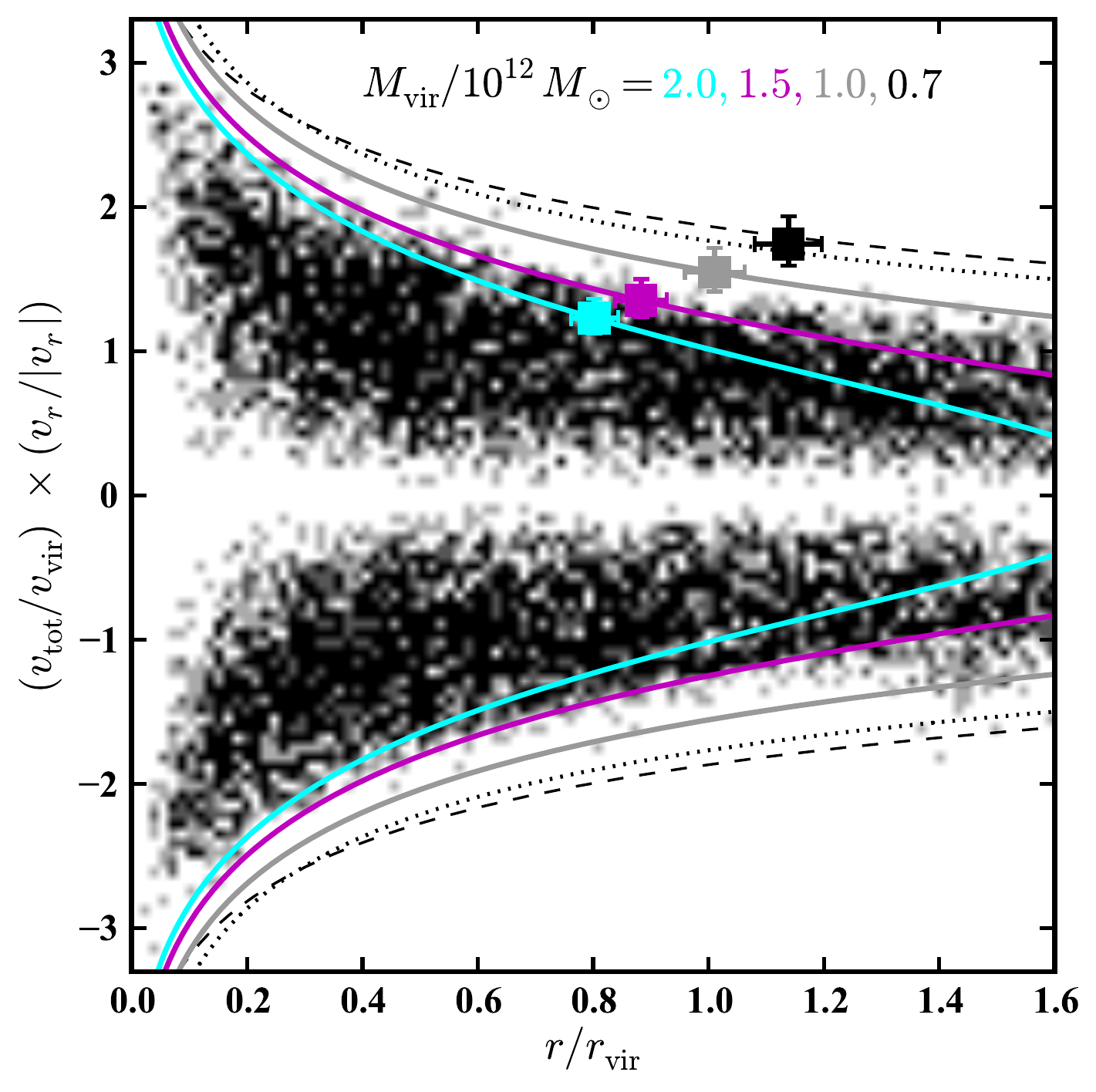}
 \caption{The phase space diagram for Aquarius subhalos using their space
   velocity; velocities and positions are scaled by virial values. At each
   radius, the velocity distribution is roughly Maxwellian 
   rather than Gaussian; the distributions are maximized near $\vtot \approx
   \vvir$ for a wide range of radii. The colored squares show Leo I's location
   in this phase space, using the measurement of $V_t$ from Paper I, for
   representative values of $M_{\rm vir, MW}$ (the colors are the same as in
   Fig.~\ref{fig:vr}). Colored lines show surfaces of constant energy in an NFW
   potential, while the dashed (dotted) curve shows the escape velocity for an
   NFW potential with a concentration of 8 (16). For virial masses less than
   $\approx 1.2\times 10^{12}\,\msun$, Leo I is less bound than virtually all
   Aquarius subhalos.
 \label{fig:vtot_pixel}
}
\end{figure}
The analysis of section~\ref{subsec:vr} does not require any knowledge of the
transverse velocity of Leo I. In this section, we incorporate the measurement of
$V_{\rm t, LeoI}$ from Paper I to compare the full space velocity of Leo I
to simulated \lcdm\ satellites. In Figure~\ref{fig:vtot_pixel}, we again plot
the phase space diagram of Aquarius subhalos, this time using total velocity $V$
rather than radial velocity on the vertical axis. The direction of the radial
velocity of each satellite is indicated by the sign of the total velocity:
negative (positive) for inward (outward) radial velocities.

We place Leo I on the plot for the same four choices of the Milky Way's virial
mass used in Figure~\ref{fig:vr}: $\mmw/10^{12}\,\msun=0.7$ (black), 1.0 (gray),
1.5 (magenta), and 2.0 (cyan).  Lines of constant energy for an object with Leo
I's Galacto-centric distance and observed velocity for each of these halo masses
are also plotted using the same color scheme, assuming a dark matter halo
following a \citet*[hereafter NFW]{navarro1997} profile with a concentration of
$c=12$ \citep{maccio2007}.  Also plotted in Figure~\ref{fig:vtot_pixel} are
curves corresponding to the escape velocity $V_{\rm esc}$ as a function of
radius for an NFW profile with a concentration parameter of 8 (black dashed
curve) and 16 (black dotted curve), assuming the NFW profile is infinite in
extent. Note that while the mass of such a profile is logarithmically divergent
with radius, the gravitational potential is finite at all radii and the escape
velocity is well-defined. In this representation, these curves are independent
of the assumed halo mass. The values used ($c=8, \, 16$) approximately span the
expected $1\,\sigma$ scatter in concentration at fixed mass for Milky Way-size
halos \citep{bullock2001, maccio2007, boylan-kolchin2010, klypin2011}.

A remarkable feature of Fig.~\ref{fig:vtot_pixel} is that vanishingly few
subhalos -- to be precise, only one in our entire Aquarius sample, or 0.01\% --
have orbits that are unbound ($V > V_{\rm esc}$). Furthermore, this single
unbound subhalo is part of a massive ($\mvir \approx 1.3\times 10^{11}\,\msun$)
infalling group\footnote{This group can be seen as a deformation of the contours
  at $v_r \approx -1.5 \,\vvir$ in Figure~\ref{fig:vr}} at $r \approx
1.4\,\rvir$ in the Aquarius C halo, and is therefore a somewhat special
case. Similar results are found for VL-II (0.06 \% unbound) and GHALO (no
unbound subhalos); see the Appendix for a version of Fig~\ref{fig:vtot_pixel}
that includes VL-II and GHALO data.

All eight of these high-resolution simulations target individual Milky Way-size
dark matter halos, while the Milky Way has a close neighbor of comparable mass,
M31. In order to assess possible effects of Local Group-like environments on the
unbound fraction of dark matter subhalos, we also have analyzed a series of
Local Group simulations from the ELVIS project (Garrison-Kimmel et al., in
preparation), which includes a suite of dark matter re-simulations of Local
Group analogs in the WMAP 7 cosmology. The vast majority of ELVIS halos have no
unbound subhalos, in agreement with the Aquarius, VL-II, and GHALO results. The
$\sim 1\%$ unbound fraction in the other ELVIS halos are either (1) the result
of very recent major mergers, or (2) subhalos associated with a massive,
recently-accreted object. The first situation is not relevant for the Milky Way,
while the orbit integrations in Paper I demonstrate that the second situation is
not applicable to Leo I. The same calculations further show that Leo I has not
interacted with Andromeda over the past Hubble time, indicating that M31 has not
been a major dynamical influence on Leo I's orbit.

The negligible unbound fractions found in the ELVIS resimulations appear to be
in conflict with results from the CLUES
project,\footnote{\href{http://http://www.clues-project.org/index.html}
  {http://www.clues-project.org/index.html}} which consists of a number of
constrained simulations of the Local Group: \citet{di-cintio2012} found that
approximately 3\% of subhalos within $\rvir$ are unbound in the CLUES Local
Group analogs. This difference, however, has its origin in how escape velocities
are computed in our analysis and in \citet{di-cintio2012}, not in the properties
of the simulations themselves. The Amiga Halo Finder \citep{knollmann2009},
which was used for identifying subhalos in the CLUES runs, calculates the
gravitational potential of a halo by assuming that it is truncated at
$\rvir$. Allowing for the mass external to $\rvir$, as in our calculations, results
in larger binding energies and significantly reduces the unbound fraction. As an
example, an NFW profile with $\mvir=7\times10^{11}$ and $c=8$ places Leo I on a
parabolic orbit ($V_{\rm esc}=V_{\rm Leo I}$), whereas the same mass
distribution, truncated at $D_{\rm Leo I}$, has an escape velocity that is 20\%
lower, $\sim 160 \,\kms$. The difference between extended and truncated mass
distributions is even more pronounced for higher values of $\mvir$.

We have explicitly checked that the unbound fractions in Aquarius, VL-II, and
GHALO are comparable to those from \citet{di-cintio2012} if we artificially
truncate the mass distribution at $\rvir$. We therefore conclude that our
calculations regarding the very low unbound fractions in Milky Way-sized halos
are likely robust to the presence or absence of an Andromeda analog, so long as
the full mass distribution surrounding the halo(s) is considered. The precise
unbound fraction clearly depends on the specific definition of the truncation
radius of the gravitational potential -- or, equivalently, on the potential's
zero point. The constraints on the virial mass of the Milky Way that we derive
in Section~\ref{sec:mass} do not depend on whether Leo I is unbound in an
absolute sense (which is an ill-posed question in a cosmological context), but
only on how bound it is relative to other Milky Way satellites and to subhalos
in $N$-body simulations; these quantities are independent of the choice of
truncation radius.

\textit{Based on our analysis of the eight highest resolution \lcdm\ $N$-body
  simulations of Milky Way-sized dark matter halos performed to date, it is very
  unlikely that Leo I is on an unbound orbit.} This point is also consistent
with both theoretical models of halo formation and numerical simulations of
structure formation, as discussed in the Introduction\footnote{The same logic
  can be applied to fast-moving satellites of M31, such as And XII and XIV
  \citep{mcconnachie2012}, meaning they are also likely to be bound
  satellites.}. Although unbound orbits are quite rare in the simulations
adopted here, they are fully incorporated in our analysis. Likewise, our
analysis already includes any subhalos with energies that have been boosted by
three-body interactions. That Leo I is almost certainly bound imposes a weak
constraint on the MW mass in the context of \lcdm, however: this requires only
that $M_{\rm vir, MW} \ga 0.7 \times 10^{12}\,\msun$ (ignoring proper motion
errors). In the next section, we combine the proper motion results of Paper I
with the Aquarius subhalo data introduced above to derive more stringent lower
limits on the virial mass of the Milky Way.

\section{Constraining the Milky Way's Virial Mass}
\label{sec:mass}
While the Aquarius halos provide us with a large sample of subhalos, the host
halos themselves only give us six different virial masses. The results of
Figures~\ref{fig:vtot_pixel} and \ref{fig:vtot_vl2_ghalo} indicate that halos in
the mass range of interest for the Milky Way are very close to self-similar in
terms of the kinematics of their subhalo populations, however, when each halo is
scaled to virial quantities. We can therefore use our Aquarius sample to
interpret Leo I's motion in a variety of halo masses. In practice, this simply
means that to change from a halo with original virial quantities $(M, \, R,\,
V)$ to a halo with virial quantities $(M',\, R',\, V')$, we multiply all radii
by $R'/R$ and all velocities by $V'/V$.

Given a subhalo's current position and velocity, we compute the velocity it
would have at the Galactocentric distance of Leo I based on its binding
energy. The orbital energy of a subhalo is calculated using a spherically
symmetric NFW profile with $c=12$. We have compared this calculation to using
the full gravitational potential computed from the parent $N$-body simulation
and find very good agreement, with a small level $(< 10\%)$ of symmetric scatter
at a given radius due to non-sphericity of the gravitational potential (see,
e.g, \citealt{hayashi2007}). 

The probability that a random subhalo $i$ has a binding energy $\scripte$ less
than that of Leo I (i.e., that the subhalo is less bound than Leo I) --
$p(\scripte_i < \scripte_{\rm Leo \, I})$ -- is then equal to the probability of
subhalo $i$ having a space velocity greater than that of Leo I. We incorporate
the Monte Carlo samplings of the proper motion error space described in Paper I
when computing $p_i = p(\scripte_i < \scripte_{\rm Leo \, I})$.  For an ensemble
of $N$ subhalos, the probability that one subhalo chosen at random is less bound
than Leo I is
\begin{equation}
  \label{eq:2}
  P(\scripte < \scripte_{\rm Leo\, I}) = \frac{1}{N} \sum_i p_i\,.
\end{equation}
We consider only satellites with positive radial velocities (subhalos moving
away from the halo center, as is the case for Leo I) in our analysis in order to
make a fair comparison to the dynamics of Leo I.

\begin{figure*}
  \includegraphics[scale=0.57]{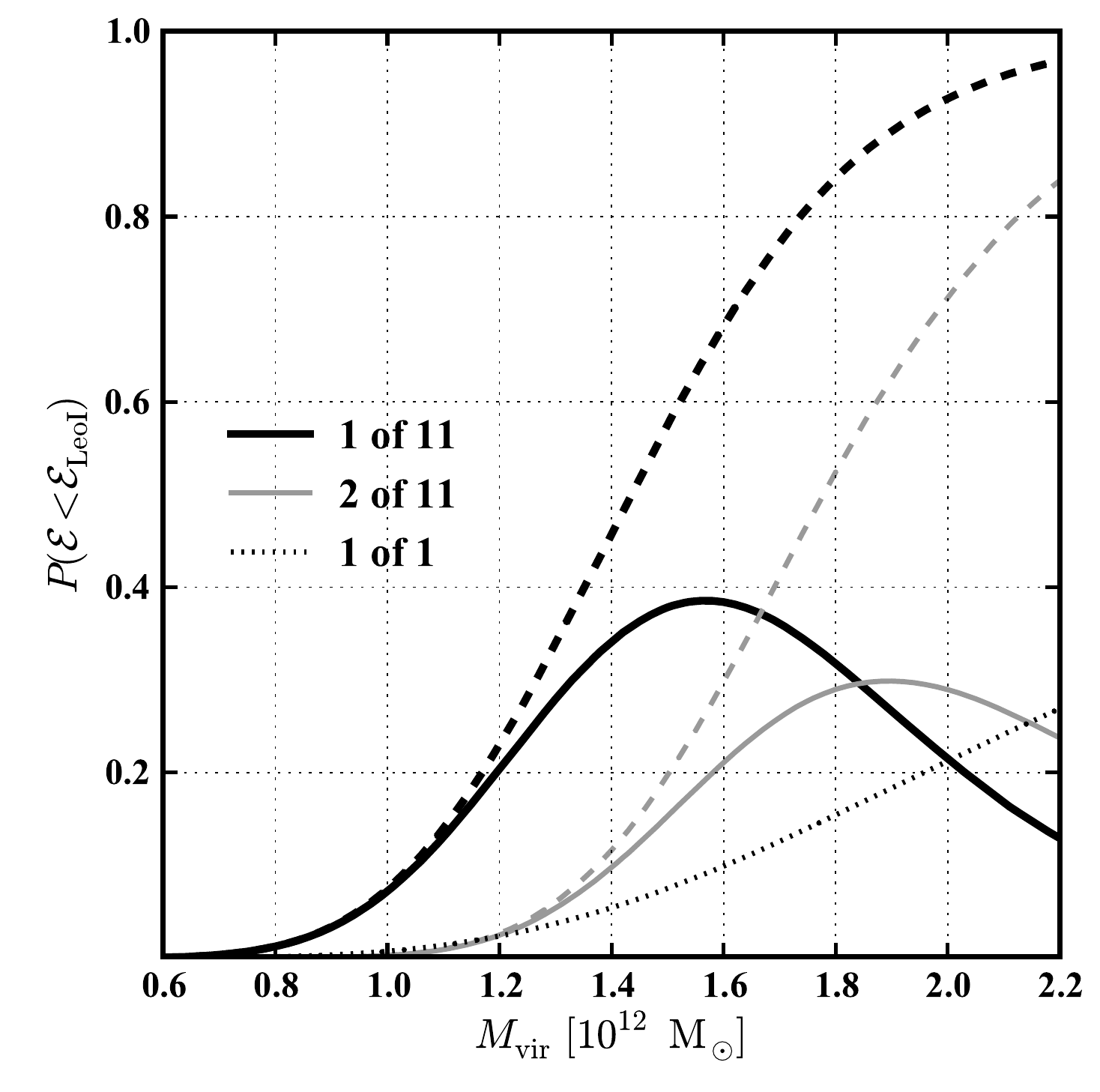}
  \includegraphics[scale=0.57]{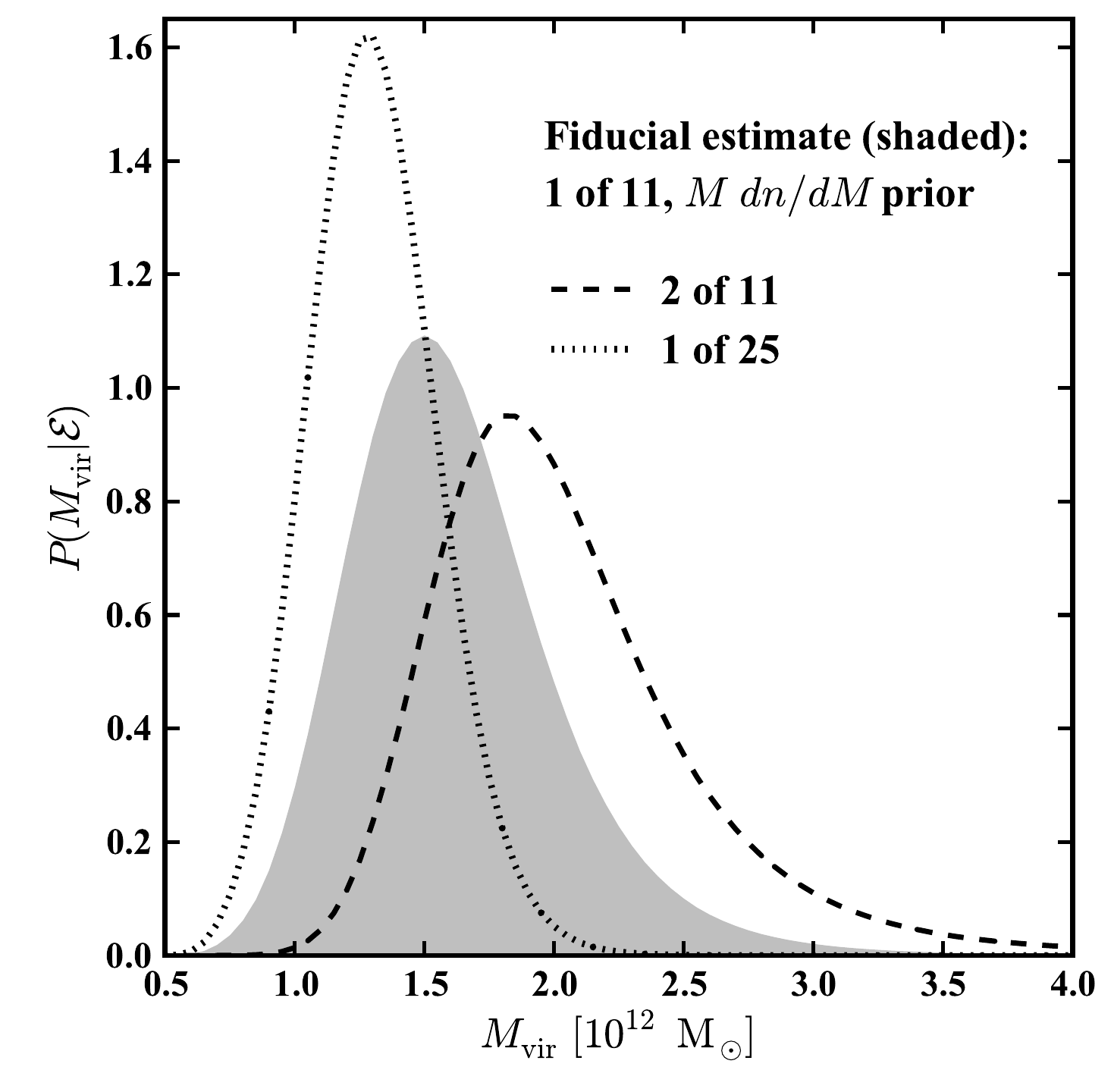}
  \caption{ \textit{Left}: Probability of finding $N$ galaxies less bound than
    Leo I as a function of the Milky Way virial mass $\mvir$ when choosing among
    $M$ galaxies with well-known energies. $(N, M)=(1, 1)$ is the probability of
    an individual galaxy being less bound than Leo I (dotted curve), while
    (1, 11) [solid black curve] and (2, 11) [solid gray curve] correspond to
    finding one or two such galaxies out of a sample of eleven. Dashed curves
    show the probability of having \textit{at least} 1 (black) or 2 (gray)
    galaxies as energetic as Leo I.  \textit{Right}: Posterior probability
    distribution (PPD) of the Milky Way's virial mass for various assumptions
    about sample size and prior. Our fiducial case (shaded) is the estimate we
    consider most reliable, with Leo I being the most energetic of a sample of
    eleven satellite galaxies and adopting a prior for the virial mass based on
    the mass-weighted dark matter halo mass function. In this case, the median
    value of the PPD is $\mvir=1.6 \times 10^{12}\,\msun$, with a 90\% confidence
    interval of $\mmw \in [1.0, 2.4] \times 10^{12}\,\msun$. If we instead assume
    that one additional MW satellite is more energetic than Leo I, we obtain a
    PPD given by the dashed curve, which is shifted to higher values of
    $\mmw$. The dotted curve shows the PPD for the case where Leo I is the most
    energetic of 25 satellites (the approximate number of known Milky Way
    satellite galaxies). Each of these PPDs assumes a prior that is proportional
    to the mass-weighted dark matter halo mass function, $\mvir\,dn/d\mvir$. A
    robust result of our analysis is that $\mmw > 10^{12}\,\msun$ at 95\%
    confidence when considering the eleven classical satellites of the Milky Way,
    irrespective of the unknown tangential velocities of some of these satellites
    and of the choice of prior.
    \label{fig:m_mw}
  }
\end{figure*}

The left panel of Figure~\ref{fig:m_mw} shows probability distributions for
finding subhalos on orbits at least as energetic as that of Leo I. The dotted
curve shows the result when considering any individual galaxy. Even at
$\mvir=2\times 10^{12}\,\msun$, a randomly selected subhalo only has a 20\%
chance of being less bound than Leo I. Of course, part of the motivation for
obtaining measurements of Leo I's proper motion was its high radial velocity,
i.e., Leo I is \textit{not} a randomly chosen satellite. The probability of
finding a high velocity satellite will obviously increase as the sample size of
satellites increases.

Ideally, we would use proper motion data for a statistically representative set
of satellite galaxies. Unfortunately, only five Milky Way satellites have
measured space velocities that are accurate at the 25\% level\footnote{These
  satellites are the Large and Small Magellanic Clouds \citep{kallivayalil2006,
    kallivayalil2006a, piatek2008}, Fornax \citep{piatek2007}, Sculptor
  \citep{piatek2006}, and Leo I \citep{sohn2012a}.}. This is still a data set
without an obvious, homogeneous selection function, although we note it does
consist of the five most luminous Milky Way satellites (excluding the
Sagittarius dwarf spheroidal). A blind search for Galactic satellites,
irrespective of luminosity, will preferentially find nearby satellites owing to
luminosity bias. This bias does not affect searches within $\sim 400\,\kpc$ of
the Milky Way for satellites with $L_V \ga 10^5 \,\lsun$, so long as there is
not a population of very extended satellites above this luminosity with surface
brightnesses that fall below current detection limits. The 11 ``classical''
satellites of the Milky Way (e.g., \citealt{mateo1998}) therefore comprise a
sample with a relatively well-known selection function. We can then ask how
likely it is to find a satellite on an orbit as energetic as Leo I's when
selecting from a sample of this size. Note that this should result in lower
limits for the MW mass that are conservative (in the sense that different
assumptions about the unknown transverse velocities of six classical satellites
will only lead to an increased lower limit on $\mmw$), as fewer than half of
these 11 satellites have measured space velocities.

The lack of strongly constraining proper motion data for 6 of the 11 classical
satellites means that, in principle, there could be as many as 6 additional
satellites on orbits as energetic as Leo I. We can estimate how likely this is
on a satellite-by-satellite case by comparing to the simulations by first
selecting subhalos with similar Galacto-centric distance and observed radial
velocity for a given satellite, then using the distribution of space velocities
of these subhalos to compute the distribution of orbital energies. This allows
us to estimate the probability that a given satellite is on an orbit more
energetic than Leo I.  This probability is very low -- less than 1\% -- in each
case. Among the five satellites with well-measured space velocities, Leo I has
the smallest binding energy. We note that using the measurements of
\citet{kallivayalil2006}, the LMC would have a binding energy that is comparable
to, or even less than, that of Leo I. However, the revised value of the LMC's
Galacto-centric velocity (Kallivayalil et al. 2012, submitted) places it on an
orbit that is more bound than Leo I so long as $\mmw \ga 7\times
10^{11}\,\msun$.  We therefore take the scenario in which Leo I is the least
bound of the eleven classical satellites as our fiducial case.

The solid black curve in the left panel of Figure~\ref{fig:m_mw} shows the
probability of precisely one subhalo from a randomly-selected sample of eleven
subhalos having a binding energy equal to or less than that of Leo I. For
comparison, we also show the probability of finding precisely two galaxies from
such a sample (solid gray curve).  If we ask instead how often at least one
subhalo has a binding energy equal to or less than that of Leo I, we obtain the
dashed black curve in left panel Figure~\ref{fig:m_mw}.  This figure shows that
Leo I's measured space velocity argues against a low mass for the Milky Way
(i.e., against $\mmw \la 10^{12}\,\msun$). It is vanishingly rare for a
randomly-chosen subhalo to be as energetic as Leo I if $\mmw < 10^{12}\,\msun$,
and there is less than a 10\% chance of finding at least 1 such subhalo when
selecting a random sample of 11 from hosts with $\mmw < 10^{12}\,\msun$.

\definecolor{mbk_gray}{gray}{0.75}
\begin{deluxetable}{c|cccc}
  \tabletypesize{\footnotesize}
  \tablewidth{\columnwidth}
  \tablecaption{
    Estimates of Milky Way's virial mass (in units of $10^{12}\,\msun$).
    \label{table:masses}
  } 
  \tablehead{ 
    $V_t$
    & \colhead{0} 
    & \colhead{measured} 
    & \colhead{measured} 
    & \colhead{measured} \\ 
    prior
    & \colhead{ $dn/dM$ }
    & \colhead{ flat }
    & \colhead{ $M\,dn/dM$ }
    & \colhead{ $dn/dM$ }
  }
  \startdata
  1 of 11 & 1.1 [0.7--1.6] & 1.6 [1.1--2.5] & 1.6 [1.0--2.4] & 1.5 [1.0--2.2]\\[0.025cm]
  2 of 11 & 1.3 [1.0--2.1] & 2.1 [1.4--3.4] & 2.0 [1.4--3.1] & 1.9 [1.3--2.8]\\[0.025cm]
  3 of 11 & 1.6 [1.1--2.7] & 2.5 [1.7--4.8] & 2.3 [1.6--4.1] & 2.2 [1.6--3.6]\\[0.025cm]
  4 of 11 & 1.9 [1.3--3.6] & 3.1 [2.0--6.8] & 2.8 [1.9--5.6] & 2.6 [1.8--4.6]\\[-0.275cm]
  \enddata
  \tablecomments{Column (1) indicates the number of satellites with binding
    energies as low as Leo I; in all cases, we assume the satellites are
    selected from a sample of eleven (the classical Milky Way
    satellites). Columns (2)-(5) show the median value and 90\% confidence
    interval of the posterior probability distribution for the virial mass of
    the Milky Way, in units of $10^{12} \,\msun$. Column (2) assumes that Leo I
    has zero tangential velocity ($3\,\sigma$ away from the result of Paper I),
    whereas columns (3)-(5) adopt the measured tangential velocity and its error
    distribution but use different priors on $P(\mvir)$. When using the measured
    tangential velocity, we find that $\mmw > 10^{12}\,\msun$ at 95\%
    confidence, a result that is independent of the choice of prior and the
    unknown tangential motions of some of the classical satellites. Our best
    estimate of $\mmw$ assumes that Leo I is the most energetic classical
    satellite and uses a mass prior that is proportional to the mass-weighted
    dark matter halo mass function (row one, column four), giving a median value
    of $\mmw=1.6 \times 10^{12}\,\msun$.}
\end{deluxetable}

While the left panel of Figure~\ref{fig:m_mw} shows the likelihood of finding
subhalo(s) less bound than Leo I in a halo with a given $\mvir$, a potentially
more interesting quantity is the posterior probability distribution of $\mvir$
given Leo I's orbital energy, $P(\mvir|\scripte)$. This can be easily calculated
using Bayes' theorem by combining the results of the left panel of
Figure~\ref{fig:m_mw} with the prior probability distribution $P(\mvir)$ for the
Milky Way mass.  One natural choice of prior in \lcdm\ is $P(\mvir)=dn/d\mvir$,
i.e., the probability of a given halo mass is proportional to the halo mass
function. Another logical \lcdm-based prior is $P(\mvir)=\mvir\, dn/d\mvir$,
which approximates a stellar mass weighting\footnote{This weighting assumes that
  the probability distribution of host halo masses for a randomly selected star
  is proportional to the halos' stellar content, and is perhaps the most
  reasonable prior if the Sun can be considered a randomly-selected star.}. We
also consider a flat prior on $\mvir$, $P(\mvir)={\rm constant}$.
Since the mass function decreases
monotonically with $\mvir$ and is never shallower than $dn/d\mvir \propto
\mvir^{-1.9}$ (e.g., \citealt{boylan-kolchin2009}), the priors incorporating the
mass function assign higher weights to low-mass halos than to those with high
mass and will result in lower estimates of $\mvir$ than the flat prior.

The results of these calculations are presented in Table~\ref{table:masses} and
plotted in the right panel of Figure~\ref{fig:m_mw}. We take as our fiducial
estimate the case where Leo I is the most energetic galaxy out of the sample of
the 11 classical Milky Way satellites with a prior of $\mvir\, dn/d\mvir$; this
results in a posterior probability distribution (PPD) for the Milky Way's virial
mass with a median value of $1.6 \times 10^{12}\,\msun$ and a symmetric 90\%
confidence interval of $\mmw \in [1.0, 2.4] \times 10^{12}\,\msun$. The PPD for
this choice is plotted as the shaded gray region in the right panel of
Figure~\ref{fig:m_mw}. The median and 90\% confidence interval of the PPD shift
to larger values as the number of satellites as energetic as Leo I is allowed to
increase: if one of the classical satellites is less bound than Leo I, then our
fiducial estimates would change to $\mmw=1.9 \times 10^{12}\,\msun$ and $\mmw
\in [1.4, 3.1] \times 10^{12}\,\msun$ at 90\% confidence. The choice of prior
has a weaker effect, as can be seen from comparing values in a given row of
Table~\ref{table:masses}. Even the effect of sampling from a larger number of
galaxies is modest: even if we assume that \textit{all} 25 known Milky Way
dwarfs had measured proper motions and that Leo I was still the most energetic
from this entire set, the median of the PPD would only decrease to $1.3 \times
10^{12}\,\msun$ with a 90\% confidence interval of $\mmw \in [0.9, 1.7] \times
10^{12}\,\msun$ for our fiducial prior.

While median and upper limits of the PPD vary based on the choice of input
parameters, Table~\ref{table:masses} shows that \textit{the data constrain (at
  95\% confidence) the Milky Way mass to exceed $10^{12}\,\msun$, independent of
  prior or number of additional fast-moving satellites.} This is the strongest
and most robust constraint provided by the combination of numerical simulations
and the Leo I proper motion measurement. Indeed, even if we assume that Leo I
has \textit{no} tangential motion, which is ruled out at $3\,\sigma$ by the
observations of Paper I, we still find that the median of the PPD for $\mmw$
exceeds $10^{12}\,\msun$ (see column 2 of Table~\ref{table:masses}).

\begin{figure}
 \centering
 \includegraphics[scale=0.55]{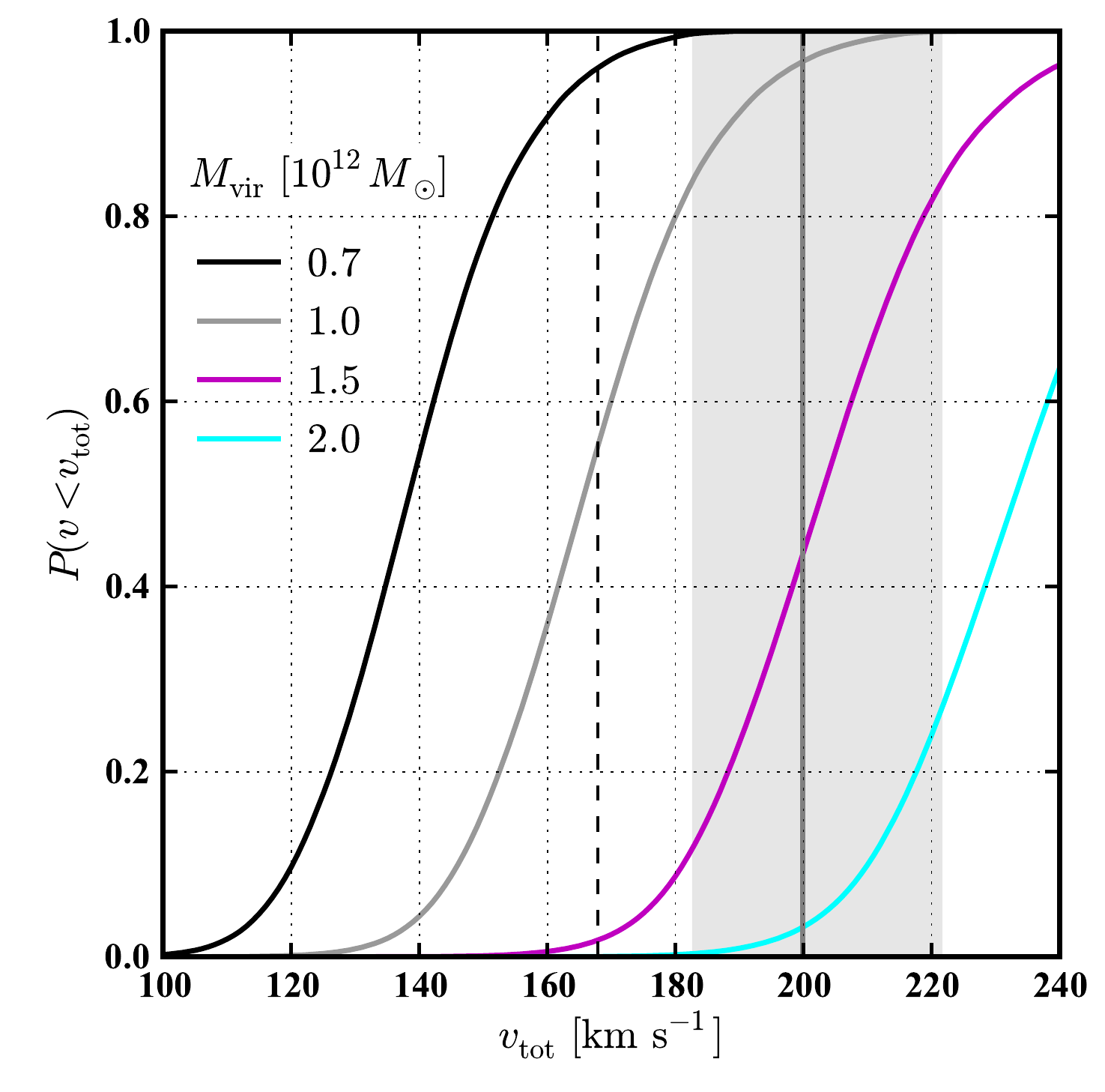}
 \caption{The probability that, when choosing from eleven subhalos, the most
   energetic will have a total velocity less than $v_{\rm tot}$ at $D_{\rm
     LeoI}$. The probabilities are plotted for four values of
   $\mmw/10^{12}\,\msun$: 0.7 (black), 1.0 (gray), 1.5 (magenta), and 2.0
   (cyan). The dashed black vertical line shows the measured $V_r$ of Leo I
   (which is an absolute lower limit for $V_{\rm LeoI}$), while the solid gray
   vertical line and gray shaded region show Leo I's measured $V_{\rm tot}$ and
   the 68.3\% confidence interval about this measurement. It is very unlikely to
   find the fastest-moving subhalo to have $V_{\rm tot}$ as high as Leo I unless
   $\mmw > 10^{12}\,\msun$.
 \label{fig:prob_v}
}
\vspace{0.2cm}
\end{figure}

The results of Figure~\ref{fig:m_mw} can also be expressed in terms of the
probability of observing an object with the velocity of Leo I for different
values of $\mmw$; this is perhaps more intuitive, as it corresponds to the
observed quantities. In Figure~\ref{fig:prob_v}, we plot the probability that
the most energetic subhalo out of a sample of 11 subhalos has a velocity (at
$D_{\rm Leo I}$) less than $V$, as a function of $V$, for four different values
of $\mmw$. The observed radial velocity of Leo I is shown as a dashed black
vertical line, while the observed total velocity of Leo I is shown as a solid
gray vertical line; the shaded gray region shows the 68\% confidence interval
about $V_{\rm tot}$. The $7 \times 10^{11}\,\msun$ Milky Way is ruled out at
95\% confidence level by the radial velocity alone (see also
Table~\ref{table:masses}); including the observed tangential velocity only
strengthens this conclusion.  In general, it is unlikely to find the most
energetic satellite (from a sample of 11) moving at Leo I's velocity unless $1
\la \mmw/10^{12}\,\msun \la 2$.

\section{Leo I and the Nature of Satellite Infall} 
\label{sec:implications}
\begin{figure}
  \centering
  \includegraphics[scale=0.53]{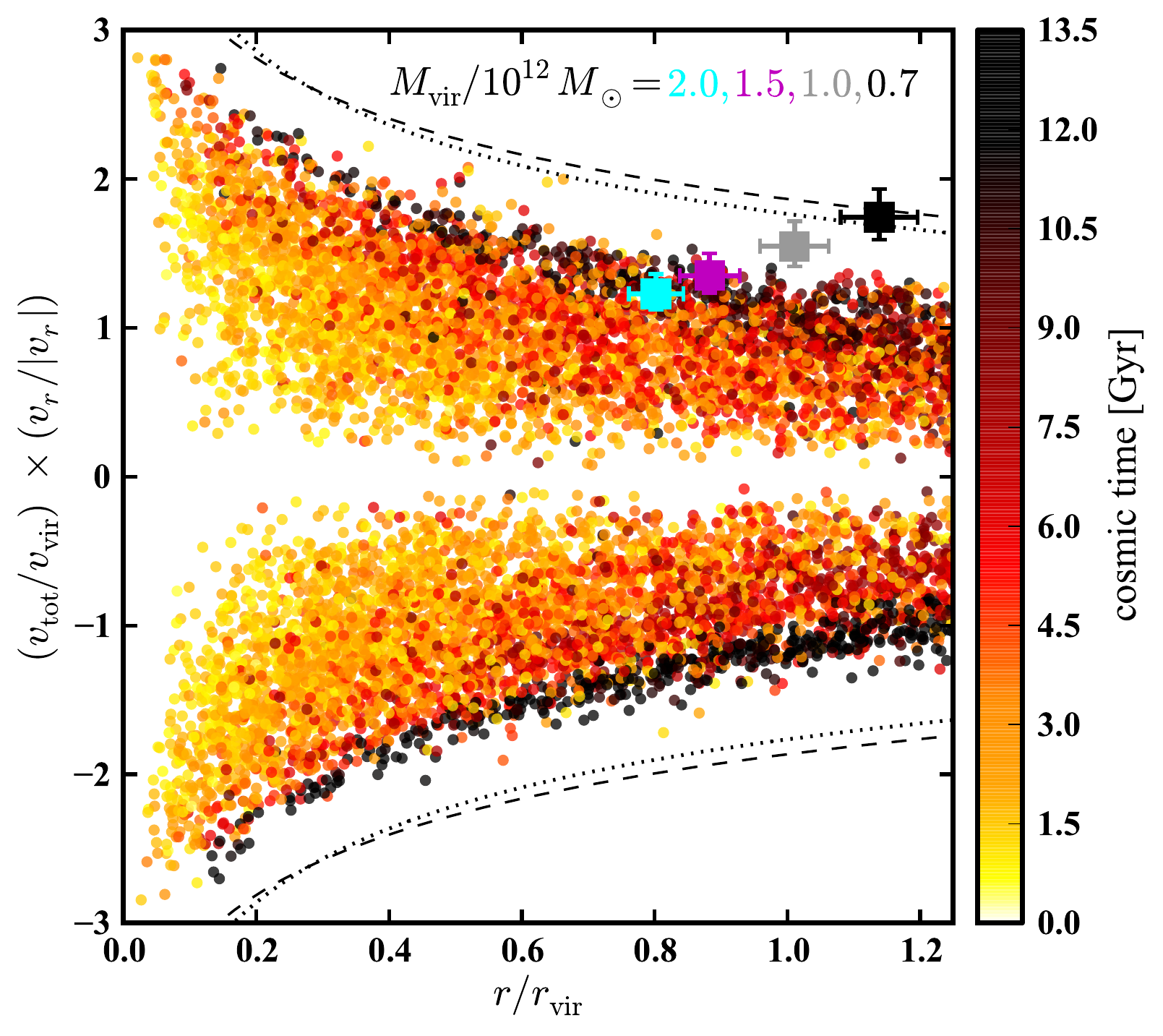}
  \caption{ Similar to Figure~\ref{fig:vtot_pixel}, except each subhalo is colored
    according to its infall time (cosmic time, in Gyr). Subhalos that fell in
    long ago (yellow-orange) exhibit a wide range of orbital energies, with the
    largest concentration of points at small radii and high binding
    energies. Recently accreted subhalos (dark red-black) occupy a much narrower
    range of orbital energies and are substantially less bound than the typical
    early-infalling subhalo. 
    \label{fig:vtot_infall_times}
  }
\end{figure}

\begin{figure*}
 \centering
 \includegraphics[scale=0.55]{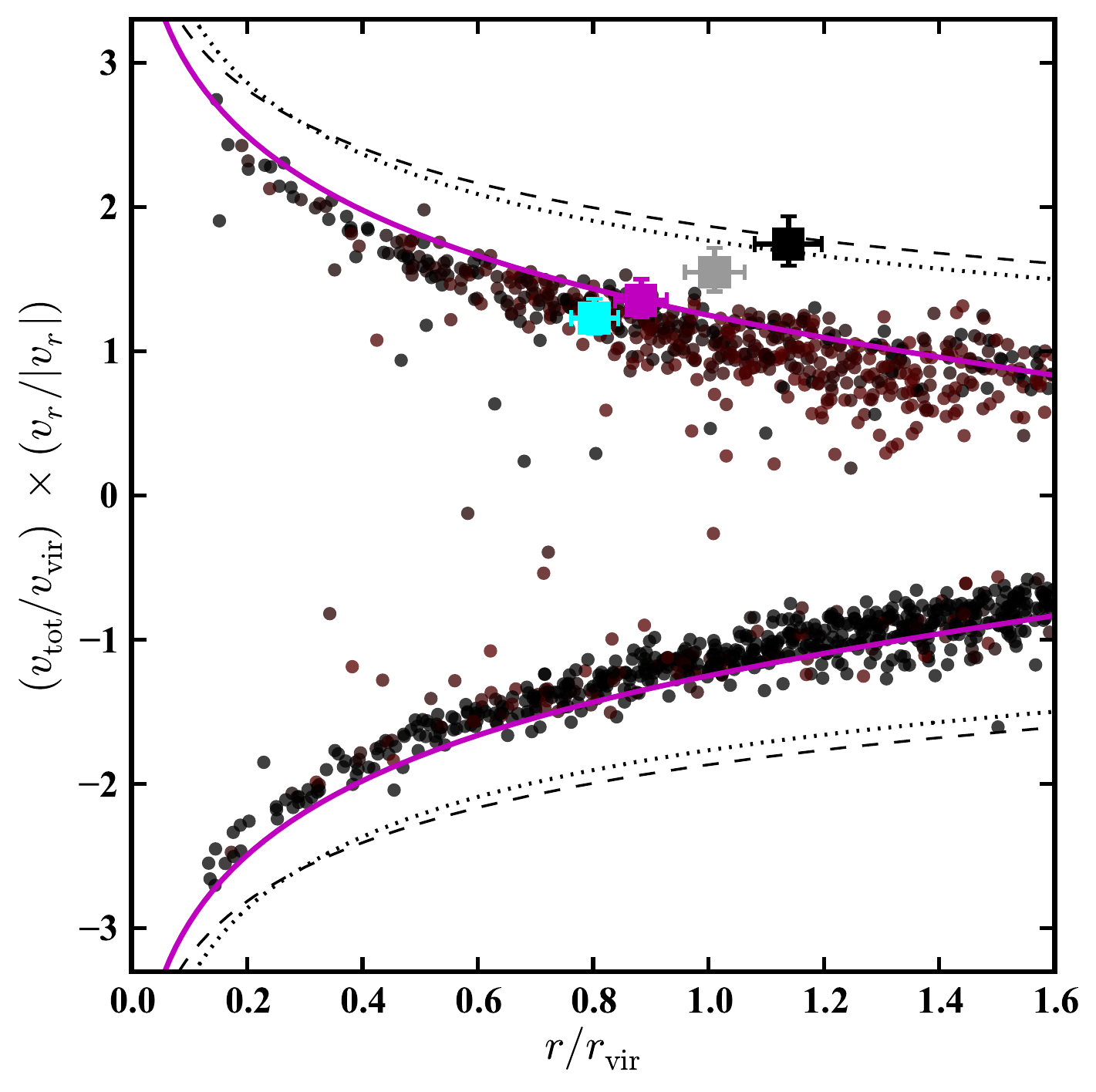}
 \includegraphics[scale=0.55, viewport=-40 0 411 410]{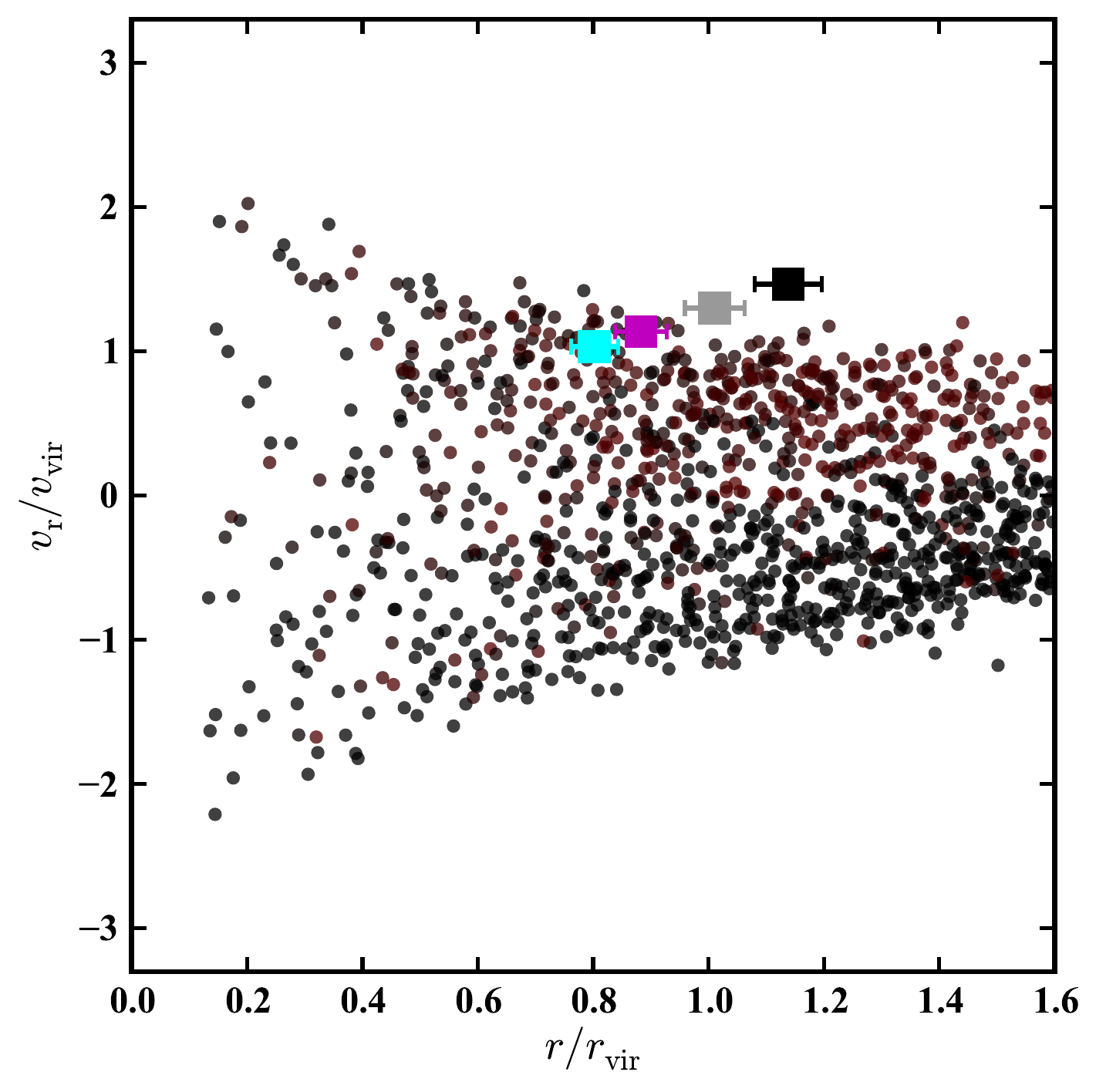}
 \caption{Phase space plots in terms of total velocity (left panel) and radial
   velocity (right panel) for subhalos accreted within the last 4 Gyr of cosmic
   time; the color scale is the same as in Figure~\ref{fig:vtot_infall_times}.
   The curve of constant energy for an object with Leo I's observed velocity and
   Galacto-centric distance in a halo of $\mmw=1.5 \times 10^{12}\,\msun$,
   $v_{\rm tot}(\rvir) \approx 1.15\,\vvir$, is shown in magenta. The dashed and
   dotted curves in the left panel are the same as in
   Figure~\ref{fig:vtot_pixel}. Whereas recently accreted satellites occupy a
   narrow range of velocities at fixed radius (left), they span a wide range of
   radial velocities (right). 
   \label{fig:vr_infall_late}
 }
\end{figure*}
 
As discussed in the Introduction, we expect to find a relationship between a
satellite's orbital energy and the time it was accreted onto its host. The
existence of such a correlation could aid in a broader interpretation of Leo I's
space velocity by constraining when Leo I fell into the Milky Way and whether or
not it has made multiple pericentric passages.

We explore the relationship between infall time and orbital energy in
Figure~\ref{fig:vtot_infall_times}, which shows a version of the total velocity
phase space for Aquarius subhalos in which each subhalo is colored according to
its infall time (measured in cosmic time, with 13.7 Gyr being the present
day). We clearly see an energy-infall time trend: while the early infalling
objects (yellow-orange) cover a wide range of orbital energies, with the
majority of subhalos being tightly bound and having apocenters of $0.75 \,\rvir$
or less, recently accreted subhalos (dark red-black) lie in a well-defined,
narrow range of energies with corresponding apocenters of $\approx 2\,\rvir$.

Recently accreted subhalos track a curve of constant energy -- corresponding to
$v_{\rm tot}(\rvir) \approx 1.15 \,\vvir$ -- quite well, especially before they
reach their first pericenters. This is fully consistent with the findings of
previous \lcdm-based analyses: \citet{wetzel2011} finds that the typical infall
velocity of satellites for hosts of $\mvir \approx 10^{12}\,\msun$ is $1.1-1.15
\,\vvir$ (see also \citealt{benson2005} and \citealt{khochfar2006}; the
excellent agreement of $v_{\rm tot}(\rvir)$ may be partially coincidental,
however, as our definition of binding energy and virial velocity differ slightly
from those of Wetzel). Figure~\ref{fig:vtot_infall_times} supports the orbit
calculations of Paper I, which strongly favor the scenario in which Leo I fell
into the MW within the last $\sim 2$ Gyr and has recently completed its first
pericentric passage (see also \citealt{rocha2012}). As noted in Paper I, this
orbital history agrees very well with the observed star formation history of Leo
I, which shows continuous star formation until $\la 1$ Gyr ago, with
enhancements at 4.5 and 2 Gyr in the past, after which star formation ceased
completely (T. Smecker-Hane et al., in preparation).

The confinement of first infall orbits to a narrow range of orbital energies
shows how valuable measurements of transverse velocities can be for interpreting
satellite dynamics. We further emphasize this point in
Figure~\ref{fig:vr_infall_late} by focusing only on recently accreted subhalos
and contrasting the resulting phase space using 3D velocities (left panel) with
the phase space using only radial velocity information (right panel).
The contrast is stark: while recently accreted subhalos occupy a well-defined
and narrow range of the 3D phase space, these satellites cover a wide range of
radial velocities at every radius.  Transverse velocities clearly add a great
deal of information that is missing in the radial phase space. This also implies
that subhalos are typically accreted with non-negligible angular momentum, and
that radial orbit approximations result in substantial information loss. A
further important point is that there are recently-accreted satellites with
small radial velocities but high tangential velocities, indicating that other
Milky Way satellites may have binding energies similar to Leo I even if they
have small measured radial velocities.

The apocentric distance for galaxies with binding energies consistent with that
of Leo I in a halo of $\mvir=1.6 \times 10^{12}\,\msun$ is $\sim 650$ kpc,
comparable to the estimate of \citet{peebles2011}. The virial radius of such a
halo is $\sim 300$ kpc, meaning this apocentric distance is similar to the
turn-around radius \citep{mamon2004}. The first apocenter of an orbit after
turn-around is expected to be at $\sim 90\%$ of the turn-around radius
in the SSIM \citep{bertschinger1985, ludlow2009}, so this large apocenter value
is consistent with theoretical expectations for a galaxy on its first infall.

The results of Section~\ref{sec:implications} have interesting implications for
satellites on first infall into their host galaxies:
Figure~\ref{fig:vtot_infall_times} shows that such satellites should populate a
narrow range of energies. For a measured distance and radial velocity of a
satellite, then, the only uncertainties are the halo mass and transverse
velocity. Leo T provides just such an example for the Milky Way: a distant
satellite ($D=407 \,\kpc$) with a low radial velocity ($v_r=-61 \,\kms$ in the
Galacto-centric frame), Leo T is the only known dwarf spheroidal / transition
object near the Milky Way with substantial HI gas content, and is therefore
almost certainly falling into the Milky Way for the first time. Our best-fitting
virial mass of $1.6 \times 10^{12}\,\msun$ leads to a prediction for the
transverse velocity of Leo T of $v_t \approx 120 \pm 20\,\kms$, i.e., Leo T's
transverse velocity should be approximately two times larger than its measured
radial velocity. Future measurements of Leo T's proper motion with \textit{HST}
(Program GO-12914; PI: T. Do) will verify or disprove this prediction. Proper
motion measurements for other distant satellites such as Leo II ($D=235\,\kpc$)
and Canes Venatici I ($D=218\,\kpc$) would also be of great
interest.\footnote{\citet{lepine2011} measured a transverse velocity of $265.2
  \pm 129.4\,\kms$ for Leo II; the central value is tantalizingly large, but is
  also consistent with zero at $2\,\sigma$.}

\section{Discussion}
\label{sec:conclusion}
While the radial velocity of Leo I has been the basis of a substantial body of
work related to the mass of the Milky Way and the properties of its satellites,
our measurement of the proper motion of Leo I (in Paper I) adds vital
information about Leo I's orbit.  We have shown that it is \textit{a priori}
extremely unlikely for Leo I to be unbound to the Milky Way in \lcdm, as
vanishingly few subhalos have velocities exceeding the local escape velocity of
their host in all of the $N$-body simulations analyzed here. By itself, this is
not strongly constraining. However, we have shown that the phase space structure
of subhalos in numerical simulations is very regular. We have therefore coupled
the proper motion measurement with the simulations and have presented a new
method for combining these data sets to derive constraints on $\mmw$.

Our best estimate for $\mmw$ is $1.6\times 10^{12}\,\msun$ with a 90\%
confidence interval of $[1.0-2.4]\times 10^{12} \,\msun$. While the precise
central value and upper limit of this range depend somewhat on input
assumptions, our strongest finding is $\mmw > 10^{12}\,\msun$ at 95\%
confidence, independent of mass priors or kinematic data from additional
satellites.  Compared to previous determinations of $\mmw$ based on the Leo I
timing argument, this best-estimate value and range are similar to
\citet{zaritsky1989} and lower than, but consistent with, \citet{li2008} and
\citet{sohn2012a}.  \citet{van-der-marel2012} have computed $\mmw$ based on the
timing argument for the MW-M31 pair and find an average for $\mmw$ of
$1.63\times 10^{12}\,\msun$, in good agreement with our results.

Our results disfavor ``light'' Milky Way models, wherein $\mmw \la
10^{12}\,\msun$. An appealing feature of such models is that they help mitigate
issues \citep{klypin1999, moore1999, boylan-kolchin2011a, boylan-kolchin2012} in
reproducing observations of the Milky Way's satellites in the context of \lcdm\
\citep{wang2012, zolotov2012, vera-ciro2013, starkenburg2013}. Our work,
however, shows that it is extremely difficult to reproduce the observed velocity
of Leo I in these low mass models. While mass constraints based on tracer stars
in the Galactic halo have traditionally favored masses of $\sim 10^{12}$, we
note that an NFW halo with $\mvir=1.6\times 10^{12}\,\msun$ and $c=12$ has
$M(<150\,\kpc)=10^{12}\,\msun$, consistent with the BHB star constraint from
\citet{deason2012a}.

The results of this paper provide further motivation for measuring proper
motions for all of the classical Galactic satellites: Table~\ref{table:masses}
indicates that if just one more of these galaxies is found to have a binding
energy as low as that of Leo I, then the 95\% confidence value for the lower
limit of $\mmw$ would increase to $\approx 1.3 \times 10^{12}\,\msun$. Reducing
the proper motion errors for Leo II is an especially high priority because the
current mean value of its space motion \citep{lepine2011} is very large
($266\,\kms$ at a distance of $235\,\kpc$, comparable to that of Leo I). More
simulations are also needed to understand potential systematics of the method
presented here. In particular, simulations of Local Group -- rather than Milky
Way -- analogs are needed if we are to truly capture the nature of the Milky Way
satellite system.

In addition to constraining the mass of the Milky Way's dark matter halo, the
orbital analysis of Paper I and the \lcdm\ simulations studied here both argue
for a first infall scenario in which Leo I has only recently joined the Milky
Way.  Coupled with (1) convincing evidence that star formation persisted in Leo
I until the past 0.5-1 Gyr \citep{smecker-hane2009}, and (2) the lack of HI
detected in Leo I ($M_{\rm HI} < 1.5 \times 10^{3}\,\msun$;
\citealt{grcevich2009}), this implies that gas in MW dwarf satellites can be
either expelled or removed on time-scales shorter than one crossing time (see
also \citealt{peebles2011}). If the gas was expelled via internal processes,
then further studies of Leo I may shed light on star formation feedback. If the
gas has been removed by ram pressure, Leo I may provide an interesting
constraint on the density of hot gas in the Milky Way halo.

A further implication of a recent Leo I infall, in conjunction with evidence
that both Magellanic Clouds are also on their first infall \citep{besla2007,
  boylan-kolchin2011}, is that the Milky Way system is a dynamic one, with
substantial late-time assembly. This active recent history at the dark matter
halo level may initially seem odd in the context of the implied quiescent merger
history for the Milky Way galaxy; however, it may simply reflect that the
Galaxy's quiescent merger history over the past several Gyr is coming to an
end. The Milky Way-M31 orbit, as inferred from recent measurement of M31's
proper motion \citep{sohn2012, van-der-marel2012a}, ensures that the Milky Way's
quiescent history has a maximum future duration of $\sim 4$ Gyr.

\vspace{0.1cm}

\section*{Acknowledgments} 
We have enjoyed fruitful discussions with Frank van den Bosch, Alis Deason,
Arianna Di Cintio, Michael Kuhlen, Aaron Ludlow, and Tammy Smecker-Hane.  We
thank the anonymous referee for helpful suggestions that led to improvements in
the paper. The Aquarius Project is part of the program of the Virgo Consortium
for cosmological simulations. We gratefully acknowledge the Aquarius, Via Lactea
II, and GHALO collaborations for giving us access to their simulation data, and
we thank Michael Kuhlen for providing us with a subhalo catalog from the GHALO
simulation and Mark Vogelsberger for providing the peculiar potential data for
the Aquarius A halo. MB-K thanks the Kavli Institute for Theoretical Physics and
the organizers of the ``First Galaxies and Faint Dwarfs'' workshop for providing
a stimulating environment during the development of this paper and acknowledges
support from the Southern California Center for Galaxy Evolution, a multi-campus
research program funded by the University of California Office of Research.  GB
acknowledges support from NASA through Hubble Fellowship grant
HST-HF-51284.01-A. This work was supported in part by the National Science
Foundation under grants AST-1009973 and PHY11-25915. Support for this work was
also provided by NASA through a grant for program GO-12270 from the Space
Telescope Science Institute (STScI), which is operated by the Association of
Universities for Research in Astronomy (AURA), Inc., under NASA contract
NAS5-26555. This research has made use of NASA's Astrophysics Data System.

\newpage
\bibliographystyle{apj}
\bibliography{draft}

\appendix
\section{Cosmological Parameter (In)Dependence}
\label{sec:appendix}
\begin{figure}
 \centering
 \includegraphics[scale=0.55]{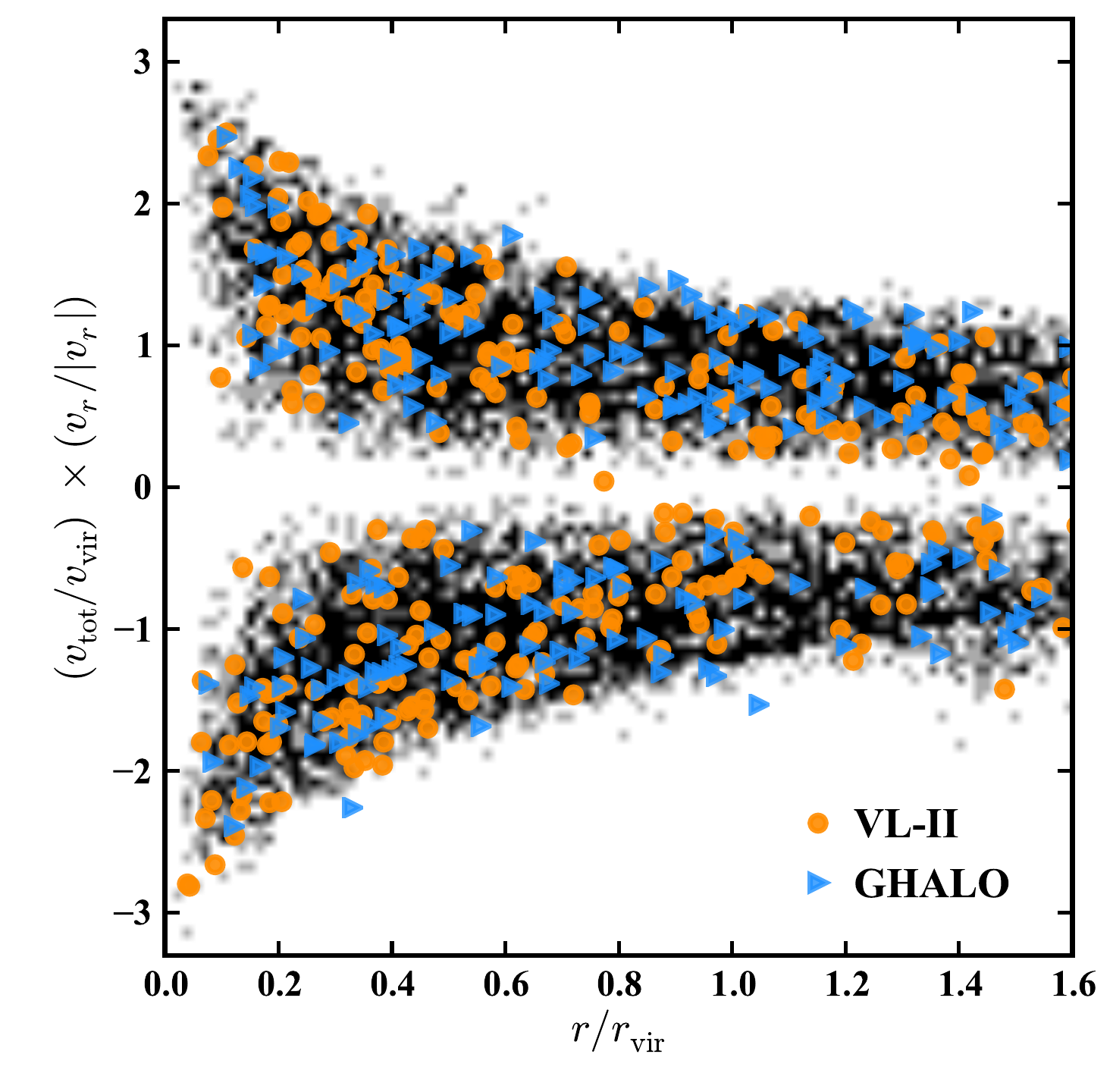}
 \caption{Same as Fig.~\ref{fig:vtot_pixel}, but also including simulations with
   WMAP3 parameters (VL-II: orange circles; GHALO: blue triangles). Subhalos
   from these simulations populate the phase space diagram in an identical
   manner to the Aquarius subhalos; our Aquarius-based results should therefore
   be robust to modest changes in cosmological parameters. Note that VL-II has a
   mass of $\mvir=1.7\times10^{12}\,\msun$, which is near the upper end of the
   Aquarius halo masses, while GHALO has $\mvir=1.08 \times 10^{12}\,\msun$,
   which is near the lower end. The virial scalings used throughout this work
   appear to be appropriate.
 \label{fig:vtot_vl2_ghalo}
}
\end{figure}
A possible concern in interpreting the orbit of Leo I through the use of the
Aquarius simulations is that the cosmological parameters adopted for Aquarius
differ slightly from the currently favored values. Specifically, WMAP7 results
indicate that $\sigma_8=0.816\pm 0.024$, $\Omega_m=0.274 \pm 0.011$, and
$n_s=0.968 \pm 0.012$ \citep{komatsu2011}, placing the Aquarius parameters
3-10\% off of the most recently measured values.  To investigate the effects of
variations in cosmological parameters, Figure~\ref{fig:vtot_vl2_ghalo}
duplicates Figure~\ref{fig:vtot_pixel} but also includes data for subhalos from
two simulations using WMAP3 parameters ($\Omega_m=0.237$, $\sigma_8=0.742$,
$n_s=0.951$): VL-II (orange circles) and GHALO (blue triangles). Together, first
and third-year WMAP parameters bracket parameters determined from the seven-year
WMAP data release. The subhalos from WMAP3-based simulations populate phase
space identically to those from the Aquarius simulations, indicating that small
changes to cosmological parameters will have no effect on interpretations of Leo
I's motion.
\end{document}

%% file: extra_preamble.tex
\newcommand{\vmax}{V_{\rm{max}}}

\newcommand{\vtot}{V_{\rm{tot}}}
\newcommand{\mvir}{M_{\rm{vir}}}

\newcommand{\rvir}{r_{\rm{vir}}}
\newcommand{\vvir}{V_{\rm{vir}}}

\newcommand{\msun}{M_{\odot}}

\newcommand{\lsun}{L_{\odot}}

\newcommand{\kpc}{{\rm kpc}}
\newcommand{\kms}{{\rm km \, s}^{-1}}

\newcommand{\zacc}{z_{\rm infall}}
\newcommand{\vacc}{V_{\rm infall}}

\newcommand{\lcdm}{$\Lambda$CDM}

\newcommand{\scripte}{\mathcal{E}}
\newcommand{\mmw}{M_{\rm vir, MW}}